\documentclass[preprint]{epl1}

\title{Depolarisation cooling of an atomic cloud}
\author{S. Hensler \and A. Greiner \and J. Stuhler \and T. Pfau}
\institute{5. Physikalisches Institut, Universit\"{a}t Stuttgart -
Pfaffenwaldring 57, 70550 Stuttgart, Germany}

\pacs{39.25.+k}{Atom manipulation (scanning probe microscopy,
laser cooling, etc.)}
\pacs{34.50.-s}{Scattering of atoms and
molecules}
\pacs{07.20.Pe}{Heat engines; heat pumps; heat pipes}

\begin{document}

\maketitle

\begin{abstract}
We propose a cooling scheme based on depolarisation of a polarised
cloud of trapped atoms. Similar to adiabatic demagnetisation, we
suggest to use the coupling between the internal spin reservoir of
the cloud and the external kinetic reservoir via dipolar
relaxation to reduce the temperature of the cloud. By optical
pumping one can cool the spin reservoir and force the cooling
process. In case of a trapped gas of dipolar chromium atoms, we
show that this cooling technique can be performed continuously and
used to approach the critical phase space density for BEC.
\end{abstract}

\section{Introduction}
Adiabatic demagnetisation~\cite{debey:1926,giauque:1927} is a well
established and very efficient cooling scheme which enables
researchers in solid state physics to cool their samples by
several orders of magnitude in a single cooling
step~\cite{deHaas:1933,giauque:19331}. However, depolarisation
processes have not yet led to a cooling concept in atomic physics.
Instead, evaporative cooling which can be observed in many fields
of physics is typically applied to obtain temperatures in the
$\un{nK}$ regime. This cooling mechanism was proposed and
demonstrated for magnetically trapped atoms by
Hess~\cite{hess:1986}. Meanwhile, it has been studied
intensively~\cite{luiten:1996,ketterle:1996} and could also
successfully be applied to atoms~\cite{ohara:2001,barrett:1995}
and molecules trapped in optical dipole traps~\cite{greiner:2003}.
Up to now, this scheme is essential to obtain degenerate atomic or
molecular quantum gases and allows nowadays to generate gases with
temperatures below $T=500\un{pK}$~\cite{leanhardt:2003}. By means
of a controllable finite trap depth $U_0$, high energetic
particles carrying more than the mean energy of a trapped particle
are allowed to escape from the trap. Rethermalisation of the
remaining particles via elastic collisions reduces the temperature
of the trapped atomic cloud and at the same time produces
particles which have sufficient energy to leave the trap again. In
typical experiments, this technique allows one to increase the
phase space density $\rho=n_0 (2 \pi \hbar^2/(m k_B T))^{3/2}$ by
several orders of magnitude, where $n_0$ and $m$ denote the peak
density and the atomic mass, respectively. The ratio
$\eta_{ev}=U_0/(k_BT)$ between the trap depth and the thermal
energy of the cloud is commonly referred as cutoff parameter. The
higher this ratio is chosen, the more energy can be carried away
by a single atom and the less atoms are lost to achieve the final
temperature. However, in this case the particles need more time to
rethermalise, so that trap losses become more significant and
finally limit the cooling process. Thus, the efficiency $\chi$ of
the cooling process is defined by the gain in phase space density
per atom loss ($-d(\ln{\rho})/d(\ln{N})$) which can be optimised
during the evaporation using the cutoff parameter. Typical
experimental values of $\chi$ range up to 4. Beside the intrinsic
high loss of atoms of this cooling method, in optical traps the
trapping volume has to be enlarged in order to reduce depth of the
trapping potential, thus the forced evaporation cooling typically
does not reach the \emph{runaway} regime where the evaporation
accelerates itself.

To reduce the temperature of an atomic sample more efficiently, we
suggest to transfer kinetic energy to an internal degree of
freedom (spin) while the cloud depolarises via inelastic dipolar
relaxation collisions. Subsequent optical pumping connects the
spin reservoir to the light field and allows to remove the energy
from the trapped atomic cloud. This cooling scheme does not rely
on removing atoms from the sample or changing the trapping
potential. Therefore, it is expected to be much more efficient
than conventional evaporative cooling.

\begin{figure}
\twofigures[scale=0.9]{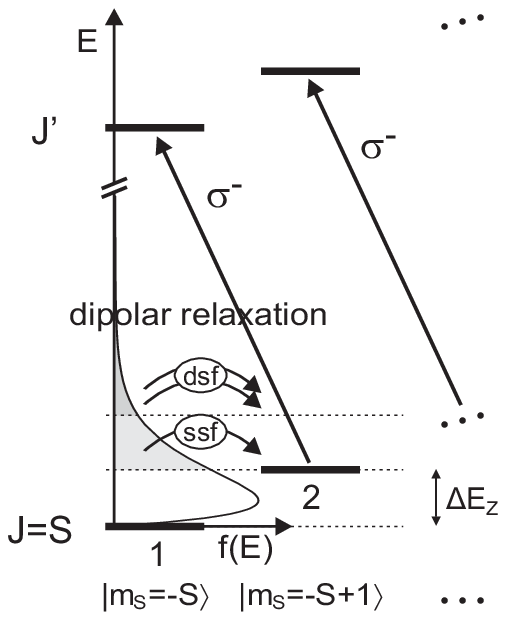}{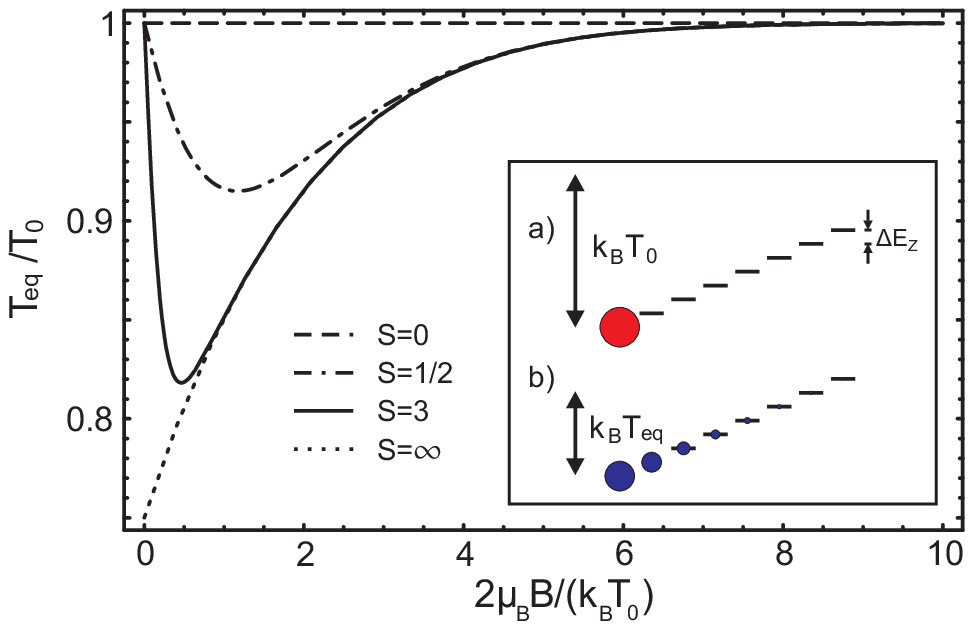} \caption{Dipolar relaxation
collisions into energetic higher state. Shown are the lowest two
energy levels of a ground and excited state manifold of a
$J\rightarrow J'=J$-transition, respectively. The levels are
separated by the Zeeman energy $\Delta E_Z$. Also indicated is the
distribution function f(E) of the relative kinetic energy of a
trapped polarised atom cloud in thermal equilibrium. Inelastic
single (ssf) and double (dsf) spin flip transitions are only
possible for atoms in the high energy tail of the distribution.
Using $\sigma^-$-polarized light the sample can be polarised in
the energetically lowest and dark state
($|m_S=-S\rangle$).}\label{f.1} \caption{Equilibrium temperature
of a previously polarised cloud depending on the magnetic field
and the spin quantum number S. The initial (a) and equilibrium (b)
situation of sample with S=3 is indicated in the inset.}
\label{f.2}
\end{figure}

In the following, we consider $N$ atoms in an homogenous magnetic
offset field $B$ trapped in a power-law potential, which is
independently of the internal state characterised by $U(x,y,z)=c_x
x^{n_1}+c_y y^{n_2}+c_z z^{n_3}$ with $\alpha=\sum_j n_j^{-1}$ and
may be realised by a far off resonant optical dipole trap. Having
in mind a specific element -- $^{52}$Cr, which has been
Bose-Einstein condensed recently \cite{griesmaier:2005} -- we
focus in this letter on atoms in a stable state without both
hyperfine structure ($I=0$) and electron orbital momentum ($L=0$),
though these are no restrictions for the cooling scheme. A finite
electron spin $S$ leads to $2S+1$ magnetic substates $|m_S\rangle$
which are energetically separated by the Zeeman energy $\Delta
E_{Z}=2\mu_B B$ (see fig.~\ref{f.1}), where $\mu_B$ is the Bohr
magneton. The dipole moment can cause inelastic dipolar relaxation
collisions in which the total spin quantum number of both atoms is
not conserved. Energy conservation requires, that for each spin
flip event to a neighbouring lower ($\Delta m_S=-1$) or higher
($\Delta m_S=1$) energetic state the energy $\Delta E_{Z}$ is
transferred to or detracted from the kinetic energy of the
colliding atoms. Moreover, we assume for the simulation, that
thermalisation between the external degrees of freedom (kinetic
reservoir) via elastic collisions occurs much faster than the
energy transfer to the spin reservoir via dipolar relaxation. Such
we can imply thermal equilibrium of the kinetic reservoir and the
total energy of the kinetic reservoir is given by
$E=(3/2+\alpha)Nk_BT$. Starting from a non-equilibrium
distribution of atoms across the states $|m_S\rangle$, the sample
gradually relaxes via dipolar relaxation to equilibrium occupation
of the states which is given by the Boltzmann distribution. For a
net relaxation rate $\dot{N}_{r}=\sum_{i=1}^{N} \dot m_{S,i}$
which contains spin flip events to neighbouring states, the
cooling rate can then be estimated using the time derivative of
the total energy of the kinetic reservoir
($\dot{E}=(3/2+\alpha)k_B(\dot{T}N+T\dot{N})$):
\begin{equation}
\label{e.1} \dot{T}=\frac{\Delta E_Z \dot{N}_{r}}{(3/2+\alpha)N
k_B},
\end{equation}
where we have neglected atom loss. Depending on the starting
condition relaxation will cool or heat the sample. The latter
process limited us to obtain Bose-Einstein condensation in a cloud
of chromium atoms in a magnetic trap~\cite{hensler:2003}.

If the atoms with temperature $T_0$ are polarised in the
energetically lowest state, transitions to the neighbouring higher
energetic substate caused by dipolar relaxation collisions cool
the sample. As the sample approaches equilibrium, the relaxation
rate and therefore the cooling rate tend to zero. The final
reachable temperature $T_{eq}$ for a certain magnetic field is
depicted in fig.~\ref{f.2} and can be calculated from:
\begin{equation}
\label{e.2} E=N\left(\frac{3}{2}+\alpha\right) k_B
T_{0}=N\left(\left(\frac{3}{2}+\alpha\right) k_B T_{eq} + \Delta
E_Z \frac{\sum_{i=0}^{2 S} e^{-\frac{\Delta E_Z}{k_B T_{eq}}
i}i}{\sum_{i=0}^{2 S} e^{-\frac{\Delta E_Z}{k_B T_{eq}}
i}}\right).
\end{equation}
Unlike in solids used for adiabatic demagnetisation where the
phonon heat capacitance in a cryogenic surrounding ($T_0\sim1 K$)
is negligible compared to the heat capacitance of the spin
reservoir, in a trapped atomic gas ($T_0\sim1-10^3\mu K$) the heat
capacitances of the spin and kinetic reservoir are of comparable
magnitude. Therefore, the achievable temperature reduction is much
smaller and an optimum ($T_{eq}/T_0=(3/2+\alpha)/(5/2+\alpha)$) is
theoretically obtained for $S \rightarrow \infty$ in the limes of
$B \rightarrow 0$. However, in atomic physics the spin reservoir
can be cooled very easily by optical pumping. In this way, this
depolarisation process can be repeated several times or even
driven in a continuous way, like we will discuss in the following.
Experimentally, polarisation of the sample in the energetically
lowest state $|m_S\rangle=-S$ can e.g. be accomplished using
$\sigma^-$-polarised light on a $J=S\rightarrow J'=J$ transition.
In this case, $|m_S\rangle=-S$ is a dark state and its population
is not affected by the pumping light (see fig.~\ref{f.1}).

In a cycle consisting of a dipolar relaxation collision and an
optical pumping transition, cooling can be provided if the Zeeman
energy $\Delta E_Z$ exceeds the energy ($E_{pol}\sim E_{rec}=k^2
\hbar^2/ (2m)$) needed to polarise the cloud. In principle this
scheme allows to generate samples with temperatures below
$T_{pol}=E_{pol}/k_B$, since in a thermal distribution high
energetic atoms $E\geq \Delta E_Z> E_{pol}$ will undergo dipolar
relaxation collisions and contribute to the cooling of the cloud.
In this respect, dipolar relaxation can be considered as an
evaporation out of the energetically lowest state, where the
cutoff parameter is given by $\eta_B=\Delta E_Z/(k_B T)$. Note,
the energy reduction of the sample works without thermalising
collisions. The required ingredients are a high inelastic dipolar
relaxation rate and a optical pumping transition back to the
initial state, which is a dark state for the pumping light. A
scheme based on elastic collisions (spin changing collisions) in
combination with a quadratic zeeman shift of the magnetic
substates was proposed by G. Ferrari~\cite{ferrari:2001}. In his
case, either linear or circular polarised pumping light provides
the necessary pumping mechanism including a dark state.

\section{Cooling model}
In the following, we develop a model to describe the continuously
driven cooling process and study the practicability of the scheme
for a sample of chromium atoms. Dipolar relaxation rates which
will be used in this section have been previously studied
experimentally and theoretically in our group~\cite{hensler:2003}.
There we found that the process is well described by dipole-dipole
scattering in the first Born approximation where either no
(elastic collision), one or both colliding atoms undergo a
transition to a neighbouring substate ($\Delta m_S=0,\pm1$). If
$\Delta M_S=\Delta m_{S,1}+\Delta m_{S,2}=-2,...,2$ accounts for
the total change in the spin quantum number of both atoms, $\Delta
M_S \Delta E_Z$ is the energy which is released (sign$(\Delta
M_S)<0$) or required (sign$(\Delta M_S)>0$) during such a
collision. The inelastic dipolar relaxation rate for an atom cloud
polarised in a extremal magnetic substate into the neighbouring
state is given by $\dot{N}_r=\dot{N}_{dip}=-\beta
N^2/\overline{V}$, where $\beta=\langle(\sigma_1+2
\sigma_2)v_{rel}\rangle_{therm}$ is the thermally averaged rate
constant containing single and double spin-flip transitions and
$\overline{V}$ is the mean trapping volume. Hereby we introduced
the collision cross sections for single spin-flip ($\sigma_1=\xi
S^3 (1+h(k_{f,1}/k_i))k_{f,1}/k_i$) and double spin-flip
($\sigma_2=\xi S^2 (1+h(k_{f,2}/k_i))k_{f,2}/k_i$) events with
$\xi=(\mu_0(2 \mu_B)^2m)^2/(30 \pi \hbar^4)$ and the relative
velocity $v_{rel}$, respectively. $h(x)$ includes the symmetries
of the particles and is defined in~\cite{hensler:2003}. Finally,
the factor $k_{f,\Delta M_S}/k_i$ accounts for the different
density of final states in the inelastic process and therefore
ensures that the energy conservation is fulfilled:
\begin{eqnarray}
  \frac{k_{f,\Delta M_S}}{k_i}= \left \{
                   \begin{array}{cl}
                      \sqrt{1-\frac{m \Delta M_S \Delta E_Z}{\hbar^2 k_i^2}} \qquad &  \tx{if} \quad 1 > \frac{m \Delta M_S \Delta E_Z}{\hbar^2 k_i^2}, \\
                      0 \qquad & \rm{otherwise}.
                    \end{array}
                   \right.
                   \label{e.3}
\end{eqnarray}
Thus, the energy transferred to spin reservoir via dipolar
relaxation collisions is given by $\dot{E}_{dip}=\tx{sign}(\Delta
M_S)\Delta E_Z \dot{N}_{dip}$. While $\Delta E_Z$ linearly
increases with the magnetic field, $\dot{N}_{dip}$ decreases for
atoms polarised in the energetic lowest substate
($|m_S\rangle=-S$). Therefore, the amount of transferred energy
can be optimised by the cut-off parameter $\eta_B$. In the case of
trapped chromium atoms (S=3), it is maximum for
$\eta_{B,opt}\approx1.31$. If the temperature of the cloud exceeds
the recoil temperature ($T\gg T_{rec}$), other heating mechanisms
can be neglected and we obtain a rough estimation \footnote{To
calculate $\eta_{B,opt}$ and eq.~(\ref{e.4}) we approximated
$(1+h(k_{f,2}/k_i))k_{f,\Delta
M_S}/k_i=1/2\cdot\Theta(v_{rel}-\sqrt{2 \mu_B \Delta M_S/m})$ for
atoms in the lowest state.} for the expect cooling rate:
\begin{equation}
\label{e.4}
\dot{T}(\eta_B)\approx-\frac{2}{3/2+\alpha}\sqrt{\frac{k_B}{\pi
m}}\xi S^2\left\{(1+\eta_B)S+(2+4\eta_B)e^{-\eta_B}\right\}\eta_B
e^{-\eta_B} \frac{N}{\overline{V}}T^{3/2}.
\end{equation}
Since the rate intrinsically depends on the dipolar relaxation
rate, the cooling process works especially well for dense samples
of atoms or molecules which possess high spin quantum numbers. The
sum within the curly brackets contains both spin flip transitions
whereas the contribution of the double spin flip transition is
suppressed by a factor of $e^{-\eta_B}$. Moreover, we find that
the time evolution of the temperature is closely connected via the
mean volume ($\overline{V}\propto T^\alpha$) to the form of the
trapping potential since a reduction in temperature simultaneously
leads to an increase in density. While a three dimensional
harmonic potential ($\overline{V}=(\sqrt{4 \pi k_B
T}/(\overline{\omega}\sqrt{m}))^3$, where $\overline{\omega}$ is
the mean trapping frequency) results in a linear change of the
temperature with the volume, a potential with $\alpha> 3/2$ --
like it can be realised in a \emph{dimple} trap -- enhances the
cooling process in time and yield a \emph{runaway} behaviour.

In order to obtain a more realistic model, we include atom loss
and the optical pumping process. Herby we will restrict our
following considerations to a three dimensional harmonic trapping
potential. Atom loss processes caused by background gas collisions
and three-body recombination are characterised by the rate
constants $1/\tau_{bg}$ and $L_{3b}$, respectively. We do not
expect two-body losses, if a sufficient deep potential is provided
($U_0\gg \Delta E_Z$ and $\dot{N}_{ev}\ll\dot{N}_{dip}$, where
$\dot{N}_{ev}$ is the evaporation rate out of the trap). The
evolution of the total atom number in the trap is then described
by:
\begin{equation}
\label{e.5}
\dot{N}=\dot{N}_{bg}+\dot{N}_{3b}=-\frac{1}{\tau_{bg}}N-\frac{L_{3b}}{\overline{V}^2}N^3.
\end{equation}
Depending on the loss process the kinetic energy of the cloud is
changed on an average by:
\begin{equation}
\label{e.6} \dot{E}_{loss}=3 k_B T \dot{N}_{bg}+2 k_B T
\dot{N}_{3b}.
\end{equation}
Since atoms lost in a three-body collision carry less than the
mean energy of a trapped gas, the remaining cloud heats up. To
account for polarisation effects, we consider the two energetic
lowest states ($|m_S=-S\rangle$, $|m_S=-S+1\rangle$) and assume a
negligible population in the other states if the cloud is
optically pumped. In the following, indices 1 and 2 will indicate
these states, respectively. The net exchange of atoms due to
dipolar relaxation is given by:
\begin{equation}
\label{e.7} \dot{N}_{r}=\dot{N}_{dip,1\rightarrow
2}-\dot{N}_{dip,2\rightarrow1}
\end{equation}
where $\dot{N}_{dip,1\rightarrow 2}$ and
$\dot{N}_{dip,2\rightarrow 1}$ are dipolar relaxation rates
containing collisions between the atoms in the same state and
atoms in state 1 and 2 which result in increase or decrease of
atoms in state 1, respectively~\cite{Giovanazzi}.
%
Including the pumping process, the population in state 1 and 2
evolve according to the following rate equations:
\begin{equation}
\label{e.8} \dot{N}_1=\dot{N}_{r}+\dot{N}\frac{N_1}{N}+((1
-\kappa)N_2 - p N_1)\Gamma_{sc},
\end{equation}
\begin{equation}
\label{e.9} \dot{N}_2=-\dot{N}_{r}+\dot{N}\frac{N_2}{N}-((1 -
\kappa)N_2 - p N_1)\Gamma_{sc}.
\end{equation}
Hereby, we introduced the scattering rate $\Gamma_{sc}$ and the
Clebsch-Gordan coefficient $\sqrt{\kappa}$ for a
$\sigma^-$-transition from state 2. The term $p N_1 \Gamma_{sc}$
takes into account that in the experiment the light is not
perfectly polarised. Absorption and emission of light are each
accompanied by a momentum transfer. Since the light scattering
rate is on the order of the dipolar relaxation rate which is
according to our assumptions much smaller than the elastic
collision rate, the sample thermalise between two scattering
events and the net energy transfer of an absorption spontaneous
emission cycle reads:
\begin{equation}
\label{e.10} \dot{E}_{pol}=(p N_{1}+ N_{2}) E_{rec}\Gamma_{sc}.
\end{equation}
These equations neither consider effects of quantum degeneracy nor
reabsorption of scattered light. In particular the latter will
limit the cooling process of an extremely dense sample. However,
effects of reabsorption can be reduced in a lower dimensional trap
or in the so called \emph{festina lente} regime \cite{cirac:1996},
where the trap frequencies exceed the linewidth of the pumping
transition. Thus, the change in the kinetic and potential energy
of the atomic cloud is given by:
\begin{equation}
\label{e.11} \dot{E}=\dot{E}_{dip}+\dot{E}_{loss}+\dot{E}_{pol}.
\end{equation}

\begin{figure}
\onefigure{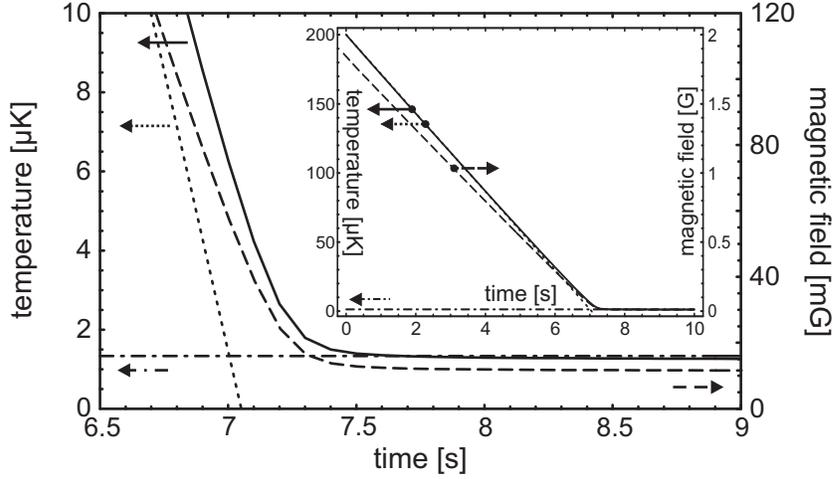} \caption{Time evolution of temperature (solid
line) and magnetic field (dashed line) during the simulated
cooling process. Additionally indicated are the solution of
eq.~\ref{e.4} (dotted line) and the temperature (dashed-dotted
line) corresponding to the energy needed to pump atoms from state
2 into state 1 by $\sigma^-$-polarised light. The same data are
shown in the inset on a different time scale.} \label{f.3}
\end{figure}
\begin{figure}
\twofigures[scale=0.7]{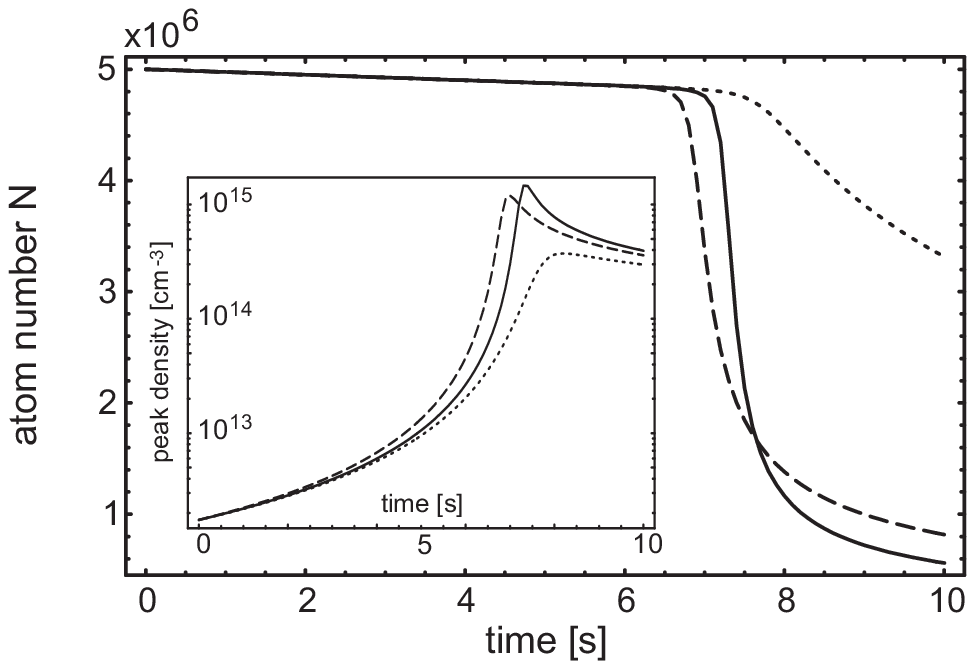}{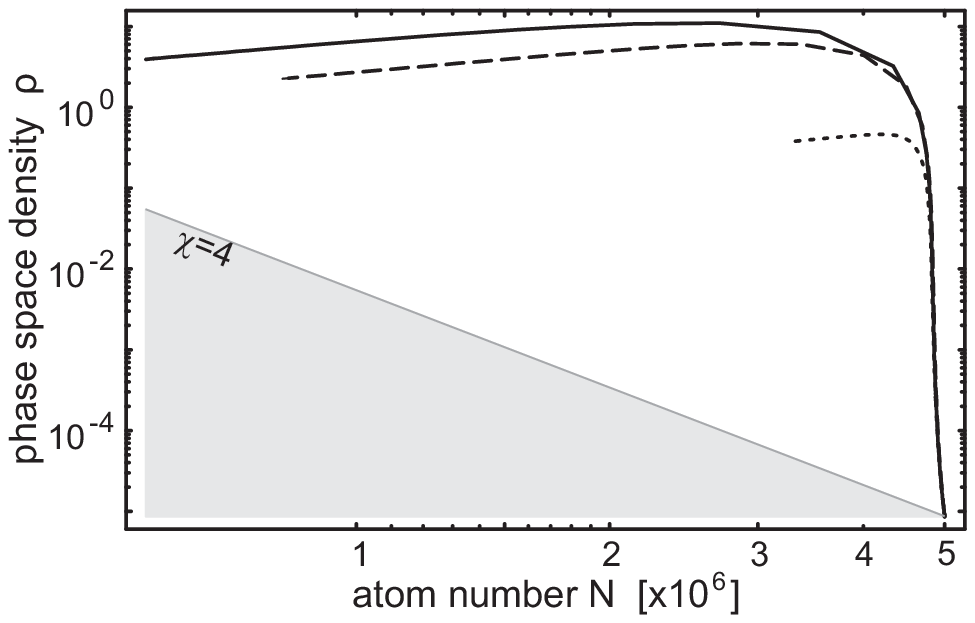} \caption{Atom number and
density (inset) dynamics during the cooling process. The solid
line correspond to the curves shown in fig.~\ref{f.3}
($p=10^{-3}$, $N_2/N_1=0.02$). The dotted line and the dashed
curves represent simulations for $p=10^{-2}$ which accounts for
the polarisation of the pumping beam and $N_2/N_1=0.005$,
respectively, while keeping others parameters fixed.} \label{f.4}
\caption{Phase space density gain per atom loss. The displayed
curves correspond to the curves illustrated in fig.~\ref{f.4}. The
efficiency $\chi$ given by the slope of the curve, reaches a
maximum value of 570 for the solid curve. Also indicated is the
area (gray) where typical evaporation cooling trajectories would
be expected.} \label{f.5}
\end{figure}

We solved the equations~(\ref{e.5}) and (\ref{e.11}) numerically
and optimised $\eta_B$ after each time step to achieve maximum
efficiency $\chi$. The calculated time evolution of the required
magnetic field and the temperature of a sample of $5\cdot 10^6$
chromium atoms with an initial temperature of $200\un{\mu K}$ in a
harmonic trapping potential ($\overline{\omega}=2\pi\cdot
500\un{Hz}$) are depicted in fig.~\ref{f.3}. Hereby, we assumed
$p=10^{-3}$ and altered the scattering rate $\Gamma_{sc}$ between
$30-2000\un{s^{-1}}$ to fix the population in state 2 to $2\%$ of
state 1 during the simulation. Since additional heating caused by
atom loss and optical pumping are negligible during the first
$7\un{s}$, both the magnetic field (dashed line) and the
temperature (solid line) reduce linearly in time (inset
fig.~\ref{f.3}). This process is well approximated by
eq.~(\ref{e.4}) which is indicated by a dotted line. During this
time, atom loss is mainly given by background gas collisions
($\tau_{bg}=200\un{s}$) (see fig.~\ref{f.4}, solid line).  The
linear decrease in temperature leads to diverging increase in both
peak density ($\propto T^{-3/2}$) (see inset fig.~\ref{f.3} (solid
line)) and phase space density ($\propto T^{-3}$), until
three-body recombination finally limits the achievable density of
the atomic cloud. Preliminary experimental measurements on a
trapped cloud of Cr atoms suggest a rate constant $L_{3b}\sim
10^{-41}\un{m^6}$, which prevents us from obtaining densities much
higher than $10^{15}\un{atoms\,m^{-3}}$ in the simulation. A clear
signature for these collisions is the more pronounced atom loss
which can be observed after $7\un{s}$.

The density still rises for a little while, until heating
processes have the same magnitude than the cooling processes. The
cooling gets inefficient and a final almost constant temperature
close to the recoil limit is reached at a magnetic field of about
$10\un{mG}$. In our specific case the final temperature is below
the temperature which corresponds to the mean pumping energy
($E_{pol}=\sum_{j=0}^\infty \kappa^j E_{rec}$) and which is
indicated by a straight line in fig.~\ref{f.3}.

In fig.~\ref{f.5} we illustrate the cooling process in a double
logarithmic phase space density - atom number plot (solid line).
The slope represents the efficiency of the cooling process as
defined before. Until three-body collisions and heating caused by
optical pumping limit the process, 5 orders of magnitude in the
phase space density could be gained while only 5 percent of the
atoms are lost. Here we obtain values of $\chi$ starting from 90
to about 570. Then the efficiency drops and finally the phase
space density decreases again.

Fig.~\ref{f.4} and fig.~\ref{f.5} contain additional curves which
result from simulations in which we changed either the degree of
polarisation of the pumping beam $p=10^{-2}$ (dotted line) or the
ratio $N_2/N_1=0.005$ (dashed line). In the first case the pumping
process is worsened and additional photons have to be scattered to
obtain the ratio $N_2/N_1=0.02$. The extra heat reduces the
cooling rate, so that density increases slower and atom loss is
less significant. The final reachable phase space density is
reduced by 1.5 orders of magnitude. Due to the improved
polarisation of the sample, the cooling rate is increased in the
second case and the density rises faster in the beginning of the
process . However, since more photons are required to maintain the
degree of polarisation of the sample the final temperature ($\sim
1.7 \mu K$) is higher and the corresponding density is lower.

\section{Conclusion}
We studied a cooling scheme which is based on depolarisation of a
polarised cloud via dipolar relaxation collisions. Our results
imply that for a cloud of optically trapped chromium atoms with
experimentally reachable starting conditions and parameters, the
cooling scheme works extremely efficient and may allow to generate
a degenerate sample. Even if this goal can not be achieved,
excellent starting condition for evaporative cooling are obtained.
The scheme may also be applied to other atomic (e.g. He, Cs) or
molecular species which exhibit large inelastic dipolar relaxation
collision rates between magnetic substates and which can be
polarised optically. Especially for molecular samples it might be
a way to increase the density be serval orders of magnitude. The
scheme allows to remove energy from a trapped sample and to
increase its density without the need of thermalising collisions.
The scheme may also be interesting for sympathetic cooling of
mixtures of species (e.g. Cr-Rb~\cite{hensler:2004}). Interspecies
collisions may transfer energy and spin, so that optical pumping
has not to be applied to the species which undergoes dipolar
relaxation.

\acknowledgments We acknowledge S. Giovanazzi and I. Vadeiko for
providing the relevant dipolar relaxation cross sections and
fruitful discussions. This work was financially supported by the
DFG (SPP1116) and the Landesstiftung BW.

\end{document}